\documentclass[a4paper,10pt]{article}
\begin{document}

\title{On Generations}
\author{Alejandro Rivero\thanks{somewhere in Kent, England. Email: \tt rivero@wigner.unizar.es}}
\maketitle
\begin{abstract}
The well known operator ordering ambiguity could motivate the
existence of generations. This possibility is explored by exploiting
the relationship between ordering and
discretization rules.
\end{abstract}

\section{Wilson fermion as analogy}

Non Commutative Geometry is not the only area knowing how to get mass terms from nowhere. Lattice
theorists learnt time ago a nice trick from Wilson. Consider the forward and backward discrete
derivatives, $\partial_+,\partial_-$ as usual
$$
\partial_+ f(x)= {f(x+\epsilon)-f(x) \over \epsilon}, \partial_- f(x)= {f(x)-f(x-\epsilon) \over \epsilon}
$$

Suitable derivatives in the lattice can be implemented as any linear combination of 
both, ${1\over a+b}(a \partial_+ + b \partial_-)$ 
Now, by noticing that $ \partial_- (f(x+\epsilon)-f(x)) =  (\partial_+ - \partial_-) f(x)$ we can rewrite
$$
{a \partial_+ + b \partial_- \over a+b}= \frac12 (\partial_+ + \partial_-) + {a-b\over a+b}  (\partial_+-\partial_-) =
  \frac12 (\partial_+ + \partial_-) + \lambda \partial_- \partial_+
$$  
where $\lambda \equiv \epsilon {a-b \over a+b}$. Here Wilson stops and we follow up: If some technical or quantum
effect avoids the limit $\lambda$ going to zero, then the system taking account of the ambiguity should be written
$
{d\over dx} f \to \partial f + \lambda \partial^2 f
$
And, if two derivatives are done, in order to account for the ambiguity we shall rewrite
$$
m {d^2\over dx^2} f \to m (\partial+ \lambda_1 \partial^2)(\partial+\lambda_2 \partial^2) f
$$

Just as a funny notation, define $\Psi_e(x)\equiv f(x), \Psi_\mu\equiv f', \Psi_\tau\equiv f''$, then:

$$
 m {d^2 f \over dx^2} = M_0  \partial^2 \Psi_e + M_1 \partial^2 \Psi_\mu + M_2 \partial^2 \Psi_\tau
$$

Note that $M_0,M_1,M_2$ contain orders of $\epsilon^0,\epsilon^1,\epsilon^2$ respectively.

\section{Introduction}

Please keep in mind that the real game is expected to be played in the 
NCG space, where some additional sheets can accomodate more structure.

Proposals to get the structure of quantum elementary objects from
the discretization of space or its regularization
are significant in the literature. But we are not aware of any
suggestion to extract from there the spectrum of elementary
fermions. This would include four flavours of particles, two chiralities
and three generations. 

We advanced this idea last year \cite{conj}, jointly with other 
conjectures of diverse quality (mostly slippery). As a piece of
the puzzle, it was already suggested that the existence of three generations
of elementary particles is the method Nature uses to parametrize
the ambiguity of quantization. 
Here some additional toy models are proposed to explore this point.
 
This letter is mainly a progress report, 
 please refer
to an hypothetical \cite{plan} for context. Definite advance will come
someday from the interplay of lattice theory and NCG; some attention
has been put on it from the lattice side \cite{japan,luscher}, but the NCG
counterpart seems to sleep since \cite{france} et al.

\section{From Bosons to Pseudofermions}
For instance,
\begin{eqnarray*}
\frac12 m {\dot x}^2&=& \frac12 m \left({f(t+\Delta)-f(t-\Delta)
  \over 2 \Delta}\right)^2 =  \\
&=&\frac14 m \mbox{ Tr} 
\pmatrix{ \left({f_+-f_-\over 2\Delta}\right)^2 &  \cr
          & \left({f_+-f_-\over 2\Delta}\right)^2 }   
=-\frac14 m \mbox{ Tr}
\pmatrix{ & {f_- - f_+ \over 2\Delta}  \cr
           {f_+-f_-\over 2\Delta} & }^2 = \\
&=&-\frac14[D,A]^2
\end{eqnarray*}
where $A=\pmatrix{f(t+\Delta) &  \cr & f(t-\Delta)}$ and 
$D=\pmatrix{&{m\over 2\Delta} \cr {m\over 2\Delta} &}$

Alternatively, we can define a vector $|\Psi>\equiv 
\pmatrix{\psi^+ \cr \psi^-}$ and to say that
$$
\frac12 m {\dot x}^2=-\frac14 <\Psi | [D,A]^2 | \Psi> 
$$
It is possible also to try to start sooner, let say from a
naive ladder $D_L \Phi(x) \equiv \Phi(x-L)$, to see it as an infinite
matrix mechanics, and then duplicate, collect and reorder vectors
to connect with the above scheme. This naive ladder is easier to
link with other formulations, say Kauffman or Costella, to name
a pair of interesting ones.
\section{From Quantization Ambiguity to generations}

Consider $\dot x_+={f(t+\Delta)-f(t)\over \Delta}$ and
         $\dot x_-={f(t)-f(t-\Delta)\over \Delta}$. See 
\cite{pathintegrals} for their relation to  
quantization ordering rules.
For generic ordering take
 $\dot x_{(\lambda\mu)}=
 \lambda \dot x_+ +  \mu \dot x_-$.

Apply the previous method. You will need to duplicate the vector space, and 
presumably to draw a more sophisticated representation $A_f$ of functions. 

For instance, let it be 
$$f_\uparrow=\frac12 \pmatrix{f(t+\Delta)+f(t) & \cr & f(t-\Delta)+ f(t)},
 f_\downarrow=\frac12 \pmatrix{f(t+\Delta)+f(t-\Delta) & \cr &2 f(t) }$$
 $M=\pmatrix{m_1& m_3 \cr m_4 & m_2}$, and
set $D=\pmatrix{& M \cr M^* &}$, $A=\pmatrix{f_\uparrow & \cr & f_\downarrow}$

Thus, putting $m_1=m_2=2 \mu/\Delta$, $m_3=m_4=2 \lambda/\Delta$,
\begin{eqnarray*}
&&[D,A]^2=\pmatrix{& M f_\downarrow - f_\uparrow M \cr 
          M^* f_\uparrow - f_\downarrow M^*  &         }^2= \\
&=&\pmatrix{
&&-\mu{f(t)-f(t-\Delta)\over\Delta}&-\lambda{ f(t+\Delta)-f(t)\over \Delta}  \cr
&&\lambda {f(t+\Delta)-f(t)\over\Delta} & \mu {f(t)-f(t-\Delta)\over\Delta} \cr
\mu{f(t)-f(t-\Delta)\over\Delta}&-\lambda{ f(t+\Delta)-f(t)\over \Delta} && \cr
\lambda {f(t+\Delta)-f(t)\over\Delta} &-\mu {f(t)-f(t-\Delta)\over\Delta}&& 
                                                                      }^2 = \\
&=&-\pmatrix{  
\mu^2 \dot x^2_- + \lambda^2 \dot x^2_+ &-2 \lambda\mu \dot x_+ \dot x_- &&  \cr
-2 \lambda\mu \dot x_+ \dot x_- & \lambda^2 \dot x^2_+ + \mu^2 \dot x^2_- && \cr
&&...&...\cr
&&...&... 
 }
\end{eqnarray*}
So that ambiguity fixing at a combination $\lambda,\mu$ can be related via 
those toys to a two generations structure with mass eigenvalues 
proportional to $\lambda\pm\mu$. Far away from the order of magnitude of
the real thing except if $\lambda\approx\mu$, which, on the other hand,
is the usual symmetric guess, $\lambda=\mu$ to get Weyl.  

We conjecture that if our Lagrangian had second order derivatives, we should
need three generations (alternatively, it could be considered to take
different ambiguity fixing in each derivation). Simultaneous fitting of first
and second derivatives is also interesting, because it suggests special
mass relationships (for instance, $\lambda-\mu\sim 1/\Delta$).

The $A$ matrix in the example shows an aspect more sensate in the basis
of mass eigenvectors, but we have preferred to keep it diagonal. 
Out of the toy model,  more serious derivations must be done, for instance
using $[D,A]= G [D,X]= [D,X] \tilde G$ with $X$ coming from the
obvious coordinate
function $f(x)=x$, and then linking the so defined derivatives $G,\tilde G$ 
with the objects
of Barrow-Majid non-commutative calculus, and so on.

\section{From Pseudofermions to infinitesimal}

Use $\Delta$ to regularize the Lagrangian. Our differential
elements are then replaced by finite differential slices. 

A discrete differential "in" a point is an affine covector instead of
a free one, which is the usual limit case. In $0+1$ it is given by two
points, that become two parallel planes in physical space. The famous\cite{g} 
"wine dance" argument from Feynman applies: rotation of one 
point is topologically equal to exchange.

Our pseudo fermions in the previous section would absorb into themselves
this property, for the discrete summation to proceed without topological
considerations. Ideally, the rule should be applied for each of the four
differentials\footnote{If you feel uncomfort identifying differentials
and particles, try instead to say "low energy modes of an 1-brane" (sorry,
the temptation was too strng, at least I have put it on a footnote. But
please remember why they were called "p"-things, the p-branes). To identify
particle fields with special choosing of coordinate fields seems too
far-fetched, but it is appealing to think of confined fields as
"angular" coordinates, without length units. So we have mixed feeling
about it.
}
 in 3+1 space in order to get four fundamental particle fields. 
Vectorial gauge fields and anti-particles must
be added following recipes in \cite{connes}, in order to establish
dualities between homology and cohomology, thus guaranteeing
the usual properties of integral calculus. This should also require some
relationship between masses of different particles, hopefully.  

An additional clue should come from the uses of Grassman variables (for
instance, in superspace formalism). Reminder that such variables were
suggested in the XIXth century as a way to formalize 
infinitesimal calculus. A Grassman variable could be presented as a
very small interval such that its square was orders of magnitude smaller,
thus zero in approximative calculations. We could expect that our fermions
had some map into Grassman variables when the continuous limit is taken
\footnote{This is touchy, as those variables in classical mechanics are
only a trick to formalize variational calculus (we simply look for
solutions $F[f+\theta]=F[f]$); their role could be different here}.

Other attemtp to link differentials and fermions is the kahler-dirac
equation.

\section{From There to the Limit?}
Field Theory is about finding the extremal of a functional $F[f]$. To
 accomplish it,
we regularize the functional introducing two scales, two units $a,\tilde a$ 
of position and momentum respectively. Then the limit is taken to remove
the scale. If $a\tilde a \to 0$, it is called classical field theory. If
$a\tilde a \to h$, being $h$ a finite constant, it is called quantum
field theory.

Point is, references to continuous and discrete worlds seem to
forgot that the
former is built from the latter. The proof of existence of classical
objects involves a regularization jointly with a jump to the limit. It
has been so since Archimedes age. But
from Wilson we know that such proofs are equivalent to the
existence of fixed points in the space of regularizations, and that
the argument can be developed by referring anything to an scale $h_0$. Just
as the limit of a series $a_n/b_n \to 0/0 $ is got by referring all its
elements to a finite constant denominator. 

The metaphysical
justification of our toys would be to come someday to claim 
that QFT is the correct
proof of existence of classical field theory (well,
it should be closer to the proof of existence of a variational calculus). 
And that, by some reason, Nature does not like to make the jump into
this limit.
 
\section*{Postlude}

Even if you has not liked anyone of the previous, I am
sure you will like this: {\it Copenhagen}, yet playing at
the Duchess theater, up the Strand. 

I was prevented about the script, but the author has made a deep research,
more careful than the usual newspaper comments about the meetings
between Bohr and Heisenberg. Better, the main plot has a lot of
deviations you will enjoy, for instance when Bohr remembers his doubts
about the spin, or when he asks for simplicity in the dialogue, because
they "must be able to explain everything to Margaritha"

\end{document}